 \documentclass[final,3p,times]{elsarticle}

\usepackage{amssymb}
\usepackage{algorithm}
\usepackage{algorithmic}
\usepackage{float}
\usepackage{makecell}
\usepackage{multirow}
\usepackage{adjustbox}
\usepackage{lscape} 
\usepackage{caption}
\usepackage{subcaption}
\usepackage{amsmath}
\usepackage{amsfonts}
\usepackage{sidenotes}
\usepackage{natbib}
\usepackage{pifont}
\usepackage{hyperref}
\usepackage{xcolor}
\usepackage{booktabs}
\usepackage{amsmath} 
\usepackage{mathrsfs}

\hypersetup{colorlinks=true,linkcolor=blue,citecolor=green,urlcolor=red}

\journal{Health Informatics}

\begin{document}

\begin{frontmatter}



\title{Navigating the Future of Healthcare HR: Agile Strategies for Overcoming Modern Challenges}


\author[1]{Syeda Aynul Karim}
	 \ead{syedaishmee19@gmail.com}

\author[1]{Md. Juniadul Islam}
 \ead{islammdjuniadul@gmail.com}


\address[1]{Department of Computer Science, American International University-Bangladesh, Dhaka-1229, Bangladesh}



\begin{abstract}
This study examines the challenges hospitals encounter in managing human resources and proposes potential solutions. It provides an overview of current HR practices in hospitals, highlighting key issues affecting recruitment, retention, and professional development of medical staff. The study further explores how these challenges impact patient outcomes and overall hospital performance. A comprehensive framework for effective human resource management is presented, outlining strategies for recruiting, retaining, training, and advancing medical professionals. This framework is informed by industry best practices and the latest research in healthcare HR management. The findings underscore that effective HR management is crucial for hospital success and offer recommendations for executives and policymakers to enhance their HR strategies.
Additionally, our project introduces a Dropbox feature to facilitate patient care. This allows patients to report their issues, enabling doctors to quickly address ailments via our app. Patients can easily identify local doctors and schedule appointments. The app will also provide emergency medical services and accept online payments, while maintaining a record of patient interactions. Both patients and doctors can file complaints through the app, ensuring appropriate follow-up actions.
\end{abstract}


\end{frontmatter}
\section{Introduction}
\label{sec:intro}
Research in various fields employs a range of methodologies to investigate phenomena, uncover insights, and contribute to knowledge. One such approach is Qualitative Research, which center’s on gathering non-numerical data to explore the underlying reasons, opinions, and motivations behind certain phenomena\cite{i1}. Utilizing methods such as interviews, focus groups, case studies, and content analysis, qualitative researchers delve deep into the intricacies of human experiences and behavior’s. This qualitative approach allows researchers to gain rich and contextualized insights into the lived experiences of individuals and the complexities of their perspectives. In contrast, Quantitative Research emphasizes the collection of numerical data and utilizes statistical analysis to draw conclusions and make generalizations\cite{i2}. Surveys, experiments, and data analysis are integral to this method, enabling researchers to quantify trends and relationships within data. The use of statistical methods helps researchers to establish patterns and correlations in large datasets, providing a broader and more structured understanding of the phenomenon under study\cite{i4}. Quantitative research is particularly valuable in studying large populations and making predictions based on the collected data\cite{i5}.
 
To enrich research endeavor’s further, Mixed-Methods Research comes into play. This approach combines qualitative and quantitative methods to provide a more comprehensive understanding of the research problem. By triangulating data from multiple sources, researchers can gain a deeper appreciation of the complexity and nuances of the research topic\cite{i5}. This blending of qualitative and quantitative approaches allows for a more holistic exploration, potentially leading to richer and more nuanced findings\cite{i3}. Literature Review or Scoping Review methodologies play a vital role in consolidating existing knowledge on specific topics. Through systematic searches, selections, and summaries of relevant literature, researchers gain insights into the evolution of research in the field, identifying gaps and informing future investigations. Literature reviews are instrumental in establishing the theoretical foundation for a study and situating it within the broader scholarly discourse. 
Case Study methodology, on the other hand, facilitates in-depth examinations of a single case or a few cases to gain a deeper understanding of a particular phenomenon\cite{i6}. 

By studying specific instances in detail, researchers can contextualize findings and understand the complexities of real-world situations. Case studies are particularly useful when researching unique or rare occurrences, and they allow researchers to explore the interplay of multiple factors that contribute to the observed outcome .In the realm of technology and systems development, Design and Development Research holds significance\cite{i7}. This methodology focuses on creating and evaluating new artifacts or systems, often in software development or user interface design contexts. Researchers involved in design and development research work iteratively, refining prototypes and solutions based on feedback and usability testing\cite{i8}. This approach allows for the practical implementation and evaluation of innovations. Action Research takes the research process a step further by actively involving researchers and stakeholders in a collaborative process. The aim is to address real-world challenges and effect practical changes. 

Engaging with the community affected by the research, action researchers play a transformative role in shaping positive outcomes. By working closely with stakeholders, including individuals, organizations, or communities, action researchers can effect change and improve practices and processes.
For establishing causal relationships between variables, Experimental Research provides a robust framework\cite{i11, i14}. By manipulating variables and comparing outcomes, researchers can determine cause-and-effect relationships, bolstering evidence-based interventions\cite{i9}. Experimental research is particularly useful in controlled settings, where researchers can isolate and manipulate specific factors to examine their effects on the outcomes of interest\cite{i8}. Survey Research plays a pivotal role in gathering data from a large number of respondents. Through questionnaires or interviews, researchers collect valuable information on trends, opinions, and characteristics, enabling insights into broad populations\cite{i13}. Surveys are commonly used to explore public opinions, consumer preferences, and employee satisfaction, among other topics. By using surveys, researchers can collect data from a diverse and representative sample, providing valuable insights into the attitudes and behavior’s of a larger population. 
\section{Literature Review}
\label{sec:Literature Review}
\textcolor{black}{The domain of usability testing is fundamentally concerned with the essential task of selecting an appropriate technique for identifying and addressing usability challenges. Recently, there has been a growing interest in reverse engineering, a method that involves deconstructing an existing product to gain insights into its structural elements and functional characteristics\cite{i6, i13}}. This increased focus is largely due to reverse engineering's capacity to reveal hidden usability issues that may not be detected through conventional task-based testing methods\cite{i12, i15, i19}. On the other hand, traditional task-based approaches have historically served as the foundation of usability testing. They offer a systematic framework for assessing user interactions with software applications, providing a structured way to evaluate their functionality and ease of use\cite{i10}. Numerous research efforts have explored the effectiveness of reverse engineering in usability evaluation. For instance, the study by khusband et al. (2018) \cite{i25} highlighted the unique benefits of this method in identifying previously overlooked usability problems that may not be easily discerned through task-based testing alone. Furthermore, S. Ghajar (2012) \cite{i31} reinforced the advantages of reverse engineering, noting its capacity to enhance understanding of the design decisions that affect user interactions. It is crucial to recognize that task-based methodologies still possess significant value and are widely utilized. Research by zhang et al. (2020) and Lanham et al. (2016) \cite{i16, i17} emphasized the importance of task-based testing in replicating real-world user scenarios, thereby serving as an effective tool for assessing the usability of software applications in practical settings. This collection of studies collectively illustrates the intricate relationship between reverse engineering and traditional task-based methods, highlighting their respective strengths and contributions to the field of usability testing.\newline

There are many models in software engineering: 

\subsection{Spiral Model}
The Spiral Model may serve as a potentially effective approach for developing the human resource management system (HRMS) within a hospital management system (HMS) under specific conditions\cite{i19}. This model is particularly suited for complex, large-scale projects, aligning well with the diverse requirements of HRMS in a healthcare environment. Its structured methodology is advantageous given the numerous components involved in HR management and the critical nature of healthcare compliance.Moreover, it is crucial to emphasize risk management throughout each phase of the project. Hospitals handle sensitive HR data and face intricate compliance challenges\cite{i20}. The iterative nature of the Spiral Model allows for the early identification and mitigation of potential risks, thereby ensuring data security and adherence to regulations\cite{i21}. Additionally, the model's inherent flexibility is beneficial in scenarios where HR needs are subject to frequent changes due to evolving HR strategies, employee transitions, or regulatory developments. The cyclical nature of the project facilitates adjustments to shifts in project scope.Engagement with stakeholders is vital in the healthcare sector. Continuous communication with HR professionals is encouraged to maintain compliance with legal and regulatory standards. The phased methodology breaks the project into manageable segments, which is particularly advantageous in healthcare, where HR functions such as recruitment, scheduling, and compensation require systematic focus\cite{i12}. Furthermore, the Spiral Model supports early prototyping, allowing for design refinement and optimization, which aids in developing an HRMS that aligns closely with the hospital's specific requirements.However, the iterative process may lead to extended development timelines and potentially elevated costs\cite{i9}. Therefore, hospitals should carefully assess their unique HRMS needs, the complexity of the project, and their capacity for a prolonged development cycle before adopting this model. Effective project management and risk assessment capabilities are essential for the successful implementation of the Spiral Model\cite{i14}.

\subsection{V-Model}
In certain scenarios, the V-Model can be a viable method for developing a Human Resource Management System (HRMS) within a Hospital Management System (HMS). Its structured methodology is well-suited to the complex nature of HRMS in healthcare, particularly for large-scale projects. This systematic approach is crucial given the many components involved in HR management and the importance of adhering to healthcare compliance regulations\cite{i26}. Emphasizing risk management at each stage is vital, as hospitals deal with sensitive HR data and complex compliance requirements.The V-Model's phased structure allows for the identification and mitigation of potential risks during specific testing phases, thereby ensuring data security and regulatory adherence\cite{i10}. Additionally, the model is beneficial in situations where HR requirements frequently change due to new HR strategies, employee movement, or evolving regulations. Its repetitive project cycle allows for smooth adjustments to changes in project scope\cite{i22}.Stakeholder involvement is crucial in the healthcare sector, and the V-Model supports continuous communication with HR professionals to ensure compliance with legal and regulatory standards\cite{i27}. By breaking tasks into manageable phases, it provides a systematic approach to various HR aspects like recruitment, scheduling, and compensation. Furthermore, the model facilitates clear documentation and validation, which is essential for designing an HRMS tailored to a hospital's specific needs\cite{i11}.However, it is important to recognize that while the V-Model may offer structured development benefits and potentially improve management of time and costs compared to the Spiral Model, hospitals should carefully evaluate their unique HRMS needs, project complexity, and timelines before adopting this approach. Successful implementation of the V-Model requires strong project management and risk assessment skills.

\subsection{Agile Model}
The Human Resource Management System (HRMS) inside a Hospital Management System (HMS) project may in fact be developed using an agile approach. Here are some explanations as to why Agile would be a good framework for this project:
 
As a result of new laws, advancements in technology, and changing patient needs, the healthcare sector is always changing. Agile enables frequent reevaluation and adjustment to changes, guaranteeing that the HRMS remains in line with the changing requirements. Agile places a strong emphasis on producing usable software in short, incremental releases\cite{i12}. For a complicated system like an HRMS within an HMS, this can be advantageous because it enables early feedback and guarantees that crucial factures are prioritized and supplied first. Agile emphasizes strong cooperation between developers and end users, which increases user involvement\cite{i23}. In a healthcare setting, involving medical staff, HR personnel, and other stakeholders in the development process can lead to a system that better meets their needs. Backlog prioritization is a technique used in agile development strategies like Scrum to make sure the most crucial features are built first\cite{i13}. This is essential for an HRMS since some functions could be more important than others. Agile can help reduce the risks connected with complex software development projects by delivering smaller, more manageable functional increments. When handling sensitive data in a hospital setting, this can be particularly curriculum. Agile approaches promote continuous integration and testing, which aids in identifying and resolving issues early in the development process\cite{i28}. For an HRMS to be reliable and secure, this is essential. Transparent Communication: Open and transparent communication between team members and stakeholders is encouraged by agile approaches. This may result in a better comprehension of the status of the project and any potential obstacles\cite{i3}. Open and transparent communication between team members and stakeholders is encouraged by agile approaches. This may result in a better comprehension of the status of the project and any potential obstacles.

\subsection{Waterfall Model}
The Waterfall model, while structured and sequential, may not be the ideal approach for developing a Human Resource Management System (HRMS) within a Hospital Management System (HMS) due to the rapidly changing nature of the healthcare sector\cite{i7}. This model follows a linear progression where each phase must be completed before proceeding to the next, which can be problematic as healthcare requirements often evolve due to regulatory changes, technological advancements, or shifts in patient care demands\cite{i29}. Furthermore, the Waterfall model typically solicits user feedback only at the end of the development cycle, potentially resulting in a misalignment between the system and the actual needs of the HR department\cite{i12}. Additionally, the model requires a comprehensive upfront specification of requirements, which can prolong the development timeline and delay the delivery of crucial HR functionalities. Once the project is underway, making adjustments to evolving needs can lead to significant delays and additional costs\cite{i24}. The lack of emphasis on early prototyping and iterations is also a limitation, especially for a complex system like an HRMS that requires timely user input\cite{i17}. Given the rapidly changing regulatory landscape in healthcare, the Waterfall approach may struggle to accommodate these shifts effectively. Testing typically occurs after the complete system is developed, making it challenging to identify and resolve earlier issues. Moreover, the intricate nature of healthcare systems increases the likelihood of discovering unexpected requirements that weren't initially captured. While the Waterfall model may suit projects with well-defined requirements, HRMS development within an HMS would benefit more from flexible, iterative methodologies like Agile.

\subsection{The Iterative and Incremental model}
The Iterative and Incremental model is a very practical approach for creating a Hospital Management System (HMS) project's Human Resource Management System (HRMS).The Iterative and Incremental methodology may be particularly appropriate for this project for the following reasons: Continuous feedback and improvement are made possible by this architecture, which is essential for creating a system as important as an HRMS. It guarantees that the system is in line with changing requirements and user needs\cite{i13}. It concentrates on delivering the system in functionally small steps\cite{i25, i27}. The HR department will immediately benefit from the early development and delivery of fundamental HR functionalities. He model is designed to accommodate changes in requirements, making it well-suited for the dynamic and evolving healthcare industry\cite{i14}. By releasing the system in smaller chunks, any possible problems or expectations that aren't aligned with stakeholder needs can be found early on and fixed. It promotes close communication between team members and between developers and end users\cite{i30}. This may result in a system that more effectively satisfies the requirements of both HR staff and other healthcare stakeholders\cite{i16}. The iterative methodology enables the prioritization of essential HR functionalities, ensuring that the most crucial elements are created first. Continuous testing and validation take place throughout the development process, assisting in the early detection and correction of faults. For an HRMS to be reliable and secure, this is essential\cite{i15}. The progressive delivery of functionality enables the deployment of crucial capabilities sooner, which helps the HR division right away.A tangible milestone is provided for each iteration, making it simpler to measure progress and show stakeholders the outcomes\cite{i17}. Alignment of Compliance and Regulations: The iterative method enables the adoption of any modifications to Compliance and Regulations that may take place during the development phase. Iterative and Incremental approach provides a structured yet flexible framework for developing the HRMS within an HMS. It allows for continuous improvement, aligns well with the needs of stakeholders, and helps manage risks effectively.

\section{Methodology}
\label{sec:methodology}
The Agile Human Resource Management System (HRMS) model utilizes Agile methodologies to create a flexible HR framework tailored to the dynamic healthcare environment. It focuses on continuous improvement cycles, ensuring responsiveness to internal and external changes.

\begin{figure}[H]
    \centering
    \includegraphics[width=0.3\textwidth]{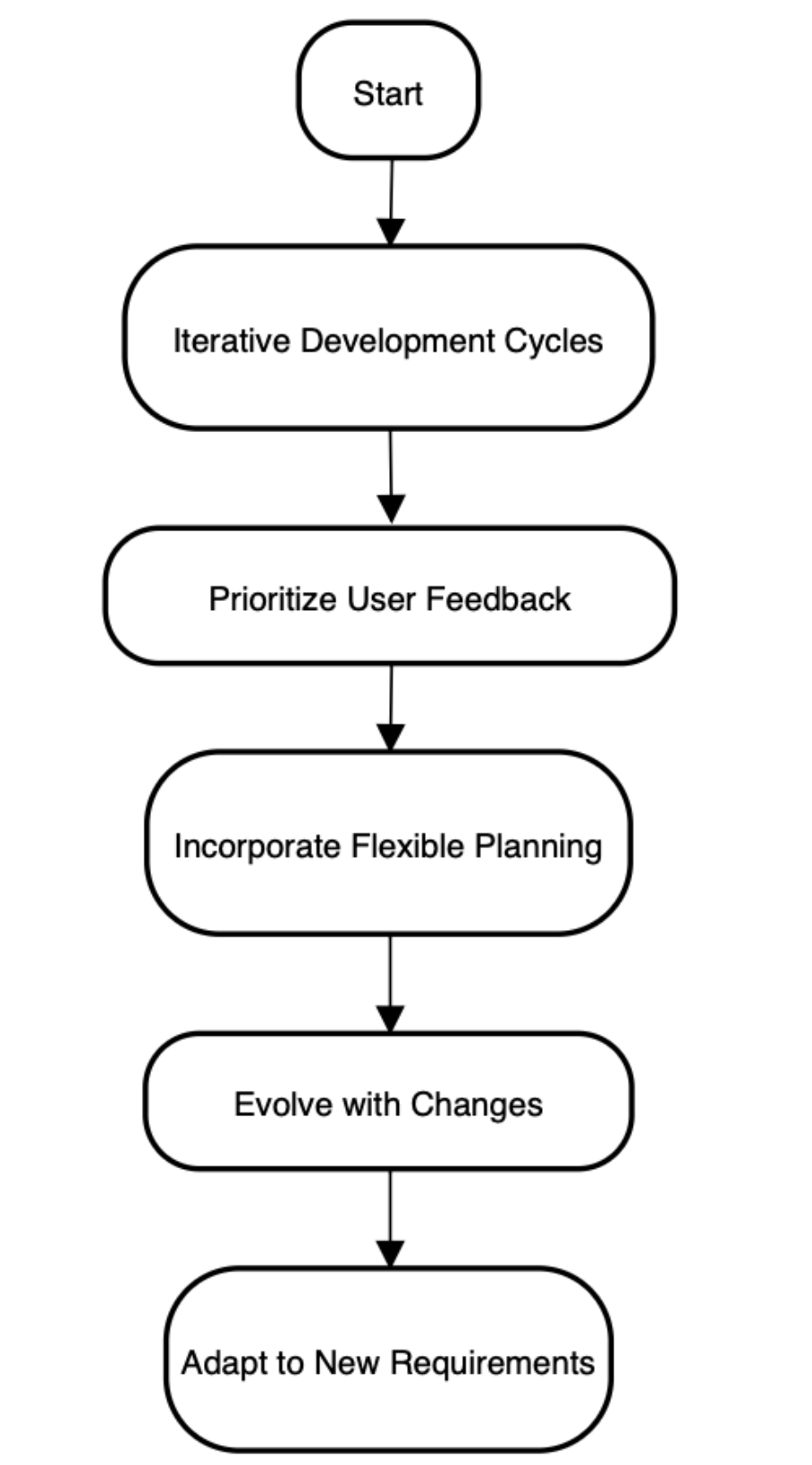}
    \caption{Flowchart of the Evaluation Process.}
    \label{fig:methodology}
\end{figure}

\hyperref[fig:methodology]{Figure~\ref*{fig:methodology} } shows some of the core applications of text to video generation that we are going to explore. These applications are briefly discussed below.

The flowchart \hyperref[fig:methodology]{Figure~\ref*{fig:methodology} } shows a visual representation of how the agile Human Resources Management System (HRMS) for guiding  work step by step in a hospital setting.

\subsection{Iterative development system}
The Agile HRMS framework starts with an iterative development process that continuously refines and improves HR tasks. Each iteration focuses on a specific HR function such as recruitment, performance management, or scheduling allowing for regular adjustments based on real-time feedback. This cycle-based approach ensures that HR processes are consistently updated to meet the evolving needs of healthcare facilities. With every cycle, key performance indicators are assessed, and necessary improvements are made. This continuous feedback loop keeps the system agile, ensuring it adapts to organizational changes and remains relevant in addressing HR challenges.

\subsection{Prioritize user feedback}
User feedback is central to the development and improvement of the Agile HRMS. The system actively collects feedback from HR personnel, medical staff, administrators, and other key stakeholders. This feedback is categorized and prioritized based on urgency and impact on operations. High-priority items, such as workflow bottlenecks or compliance issues, are addressed first. By consistently integrating user feedback into the development process, the HRMS enhances usability, satisfaction, and overall performance. This ensures that the system evolves in response to real-world needs, making it more effective for managing HR processes in dynamic healthcare environments.

\subsection{Incorporate flexible planning}
The Agile HRMS integrates flexible planning, allowing the system to adjust HR strategies in real-time based on feedback and evolving conditions. This flexibility ensures that HR managers can respond to new challenges, such as workforce shortages, changing regulations, or shifts in operational priorities without causing disruptions. Flexible planning enables proactive problem-solving and resource allocation, optimizing staffing and HR processes to ensure smooth operations. By adapting plans dynamically, the HRMS maintains alignment with both immediate organizational needs and long-term healthcare goals, ensuring operational efficiency and regulatory compliance.

\subsection{Evolve with changes}
The Agile HRMS is built to evolve continuously in response to organizational, technological, and operational changes. Feedback-driven adaptations ensure that the system addresses both immediate challenges and long-term goals. Flexible planning is embedded in the system, allowing for rapid adjustments to new conditions such as staff changes, new technologies, or policy updates. This capability ensures that the HRMS remains responsive to emerging needs and continues to optimize resource management, scheduling, and staff development. By evolving in line with real-time changes, the system ensures consistent alignment with healthcare operations.

\subsection{Adapt to regulatory changes }
In the ever-changing landscape of healthcare regulations, the Agile HRMS is equipped to handle regulatory shifts smoothly. The system monitors updates in legal standards, such as those related to patient privacy, healthcare compliance, and workforce management. When new regulations are introduced, the HRMS integrates these requirements through its flexible planning mechanisms. This ensures that the healthcare organization remains compliant without significant disruptions to HR processes. By adapting to these changes in real-time, the system minimizes risks, avoids penalties, and ensures that HR practices are aligned with legal and ethical standards.

\subsection{Adapt to operational changes}
The Agile HRMS is designed to flexibly adapt to various operational changes that may affect HR processes. Whether it's dealing with shifts in workforce dynamics, new technology implementations, or evolving organizational goals, the system adjusts its strategies in real-time. HR activities such as staff recruitment, performance evaluation, and resource allocation can be optimized quickly in response to new developments. This adaptability ensures that the HRMS remains efficient in managing day-to-day operations while addressing long-term organizational needs, helping the healthcare facility maintain a high level of productivity and service quality.

\section{Result}
\begin{figure}[H]
    \centering
    \includegraphics[width=0.7\textwidth]{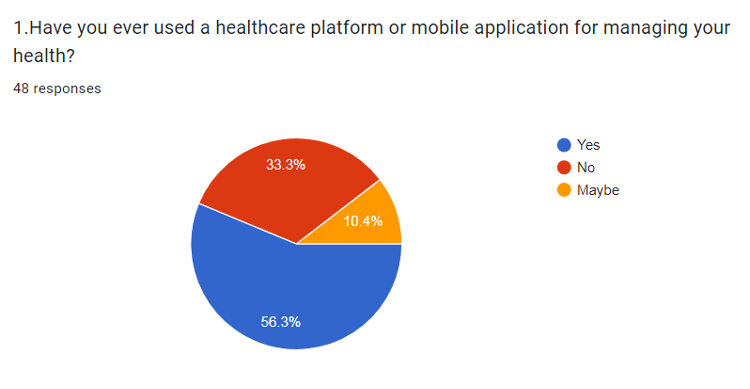}
    \label{fig:model}
\end{figure}
Mobile apps and healthcare platforms are widely used by users to manage their health. Development starts with identifying essential features and comprehending user needs while using the Agile model. These are arranged in a product backlog and given priority according to their significance. Each sprint in the development process focuses on a different set of features. Ongoing feedback loops, frequent testing, and incremental delivery guarantee that the platform efficiently satisfies user needs and changes with the times to meet changing healthcare requirements. 

\begin{figure}[H]
    \centering
    \includegraphics[width=0.7\textwidth]{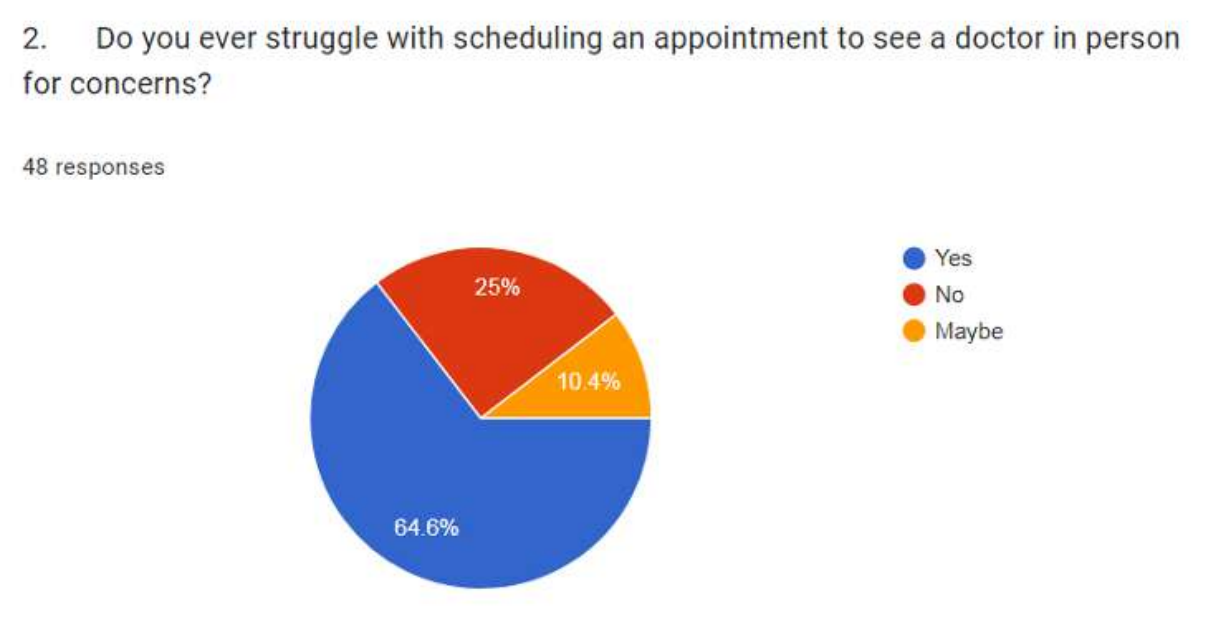}
    \label{fig:model1}
\end{figure}
Using the Agile model, a healthcare platform could incorporate a user-friendly interface for appointment scheduling. Regular feedback loops and iterative development would allow for rapid adjustments based on user testing. Continuous integration would ensure seamless integration with existing systems. Short sprints would focus on refining and optimizing the scheduling process. This iterative approach would lead to a more efficient and user centric appointment booking system, ultimately reducing the struggle for patients.

\begin{figure}[H]
    \centering
    \includegraphics[width=0.7\textwidth]{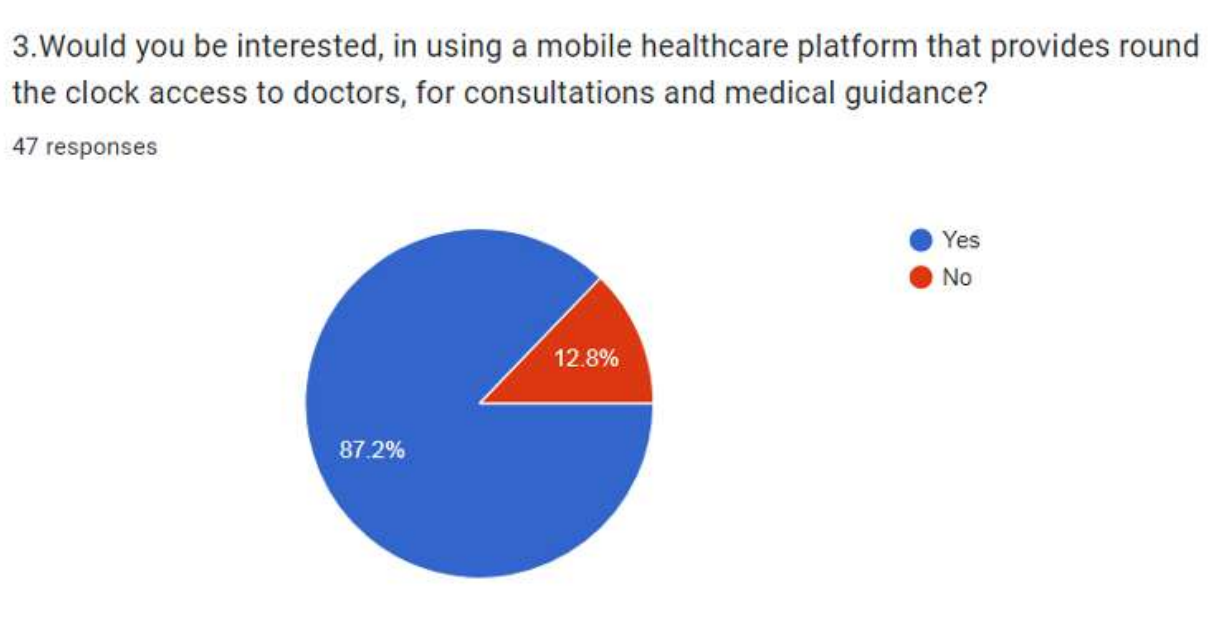}
    \label{fig:model2}
\end{figure}
Providing 24/7 access to doctors, mobile healthcare fits perfectly with the Agile model. It allows for iterative improvements, allowing for rapid adaptation to changing healthcare needs. Ongoing stakeholders ensure that the platform meets user needs and complies with medical regulations. Regular feedback and incremental updates increase user satisfaction and security. Agile’s focus on rapid value delivery can immediately benefit patients seeking medical guidance. The Agile framework effectively promotes risk mitigation and the development of a platform that can be very attractive in the rapidly evolving healthcare. 

\begin{figure}[H]
    \centering
    \includegraphics[width=0.7\textwidth]{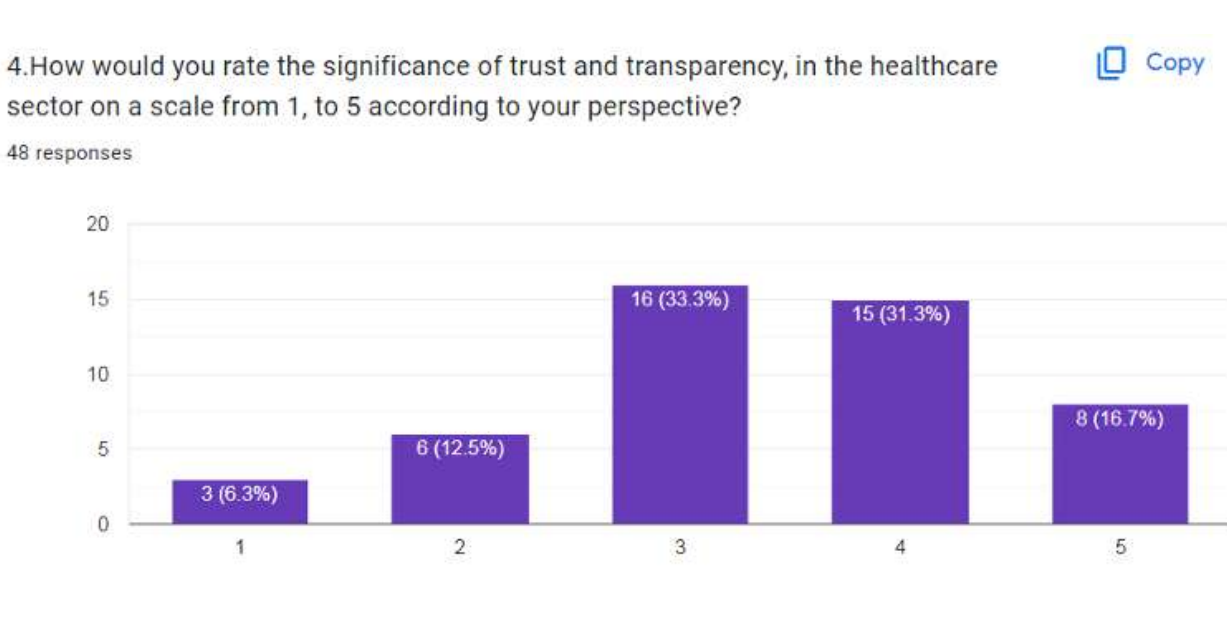}
    \label{fig:model3}
\end{figure}
In agile model the meaning of trust and straightforwardness in the medical care area would go through a ceaseless criticism circle. This iterative interaction includes standard reviews and partner conferences, guaranteeing continuous appraisal and rating of trust and transparency. Improvements would be driven by means of cooperative studios with medical care experts and innovation specialists, and key execution pointers (KPIs) would measure progress. The unique idea of the Light footed model empowers a last evaluation that considers the changing significance of straightforwardness and confidence in the medical services industry.

\begin{figure}[H]
    \centering
    \includegraphics[width=0.7\textwidth]{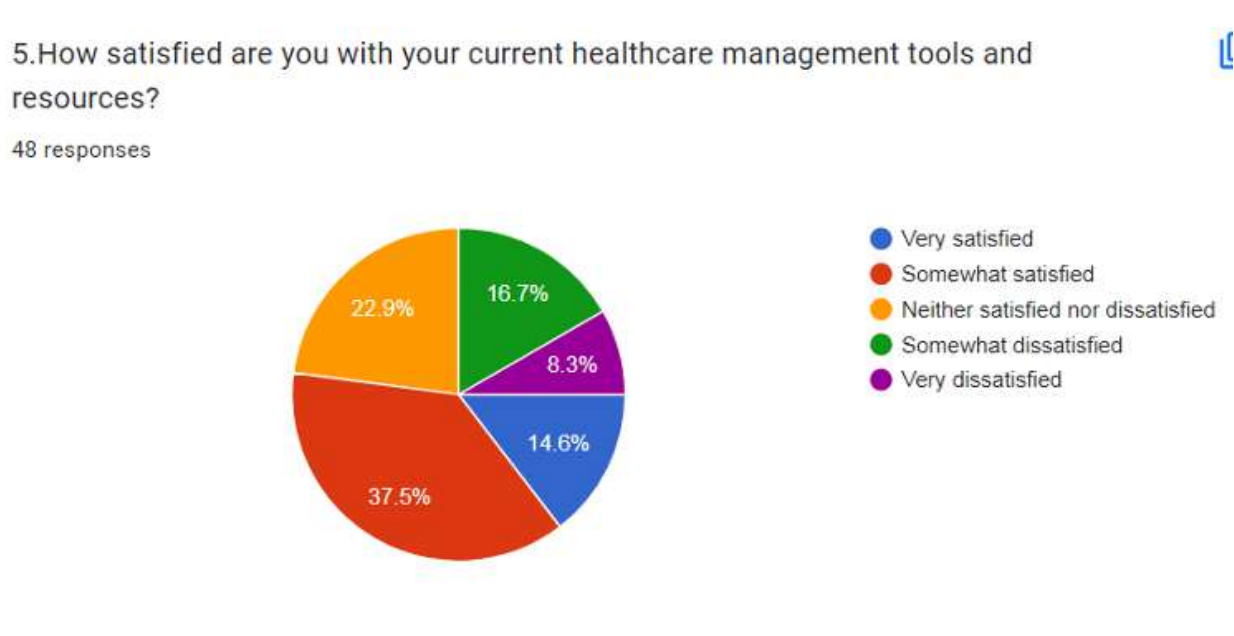}
    \label{fig:model4}
\end{figure}

The resources and instruments available for managing healthcare today frequently leave a lot of people unhappy. This dissatisfaction can be methodically addressed by using the Agile Model. User demands and feedback are regularly included into the process of improving the tools through iterative development. Brief sprints concentrate on improving features and functionality to make sure they meet user requirements. The usability and efficacy of the tools are further improved by frequent testing and feedback loops. Agile Model enables quick adjustments to shifting healthcare needs, resulting in improved management tools. In general, healthcare management resources can adapt to better suit users' requirements and expectations by putting the Agile Model into practice.

\begin{figure}[H]
    \centering
    \includegraphics[width=0.7\textwidth]{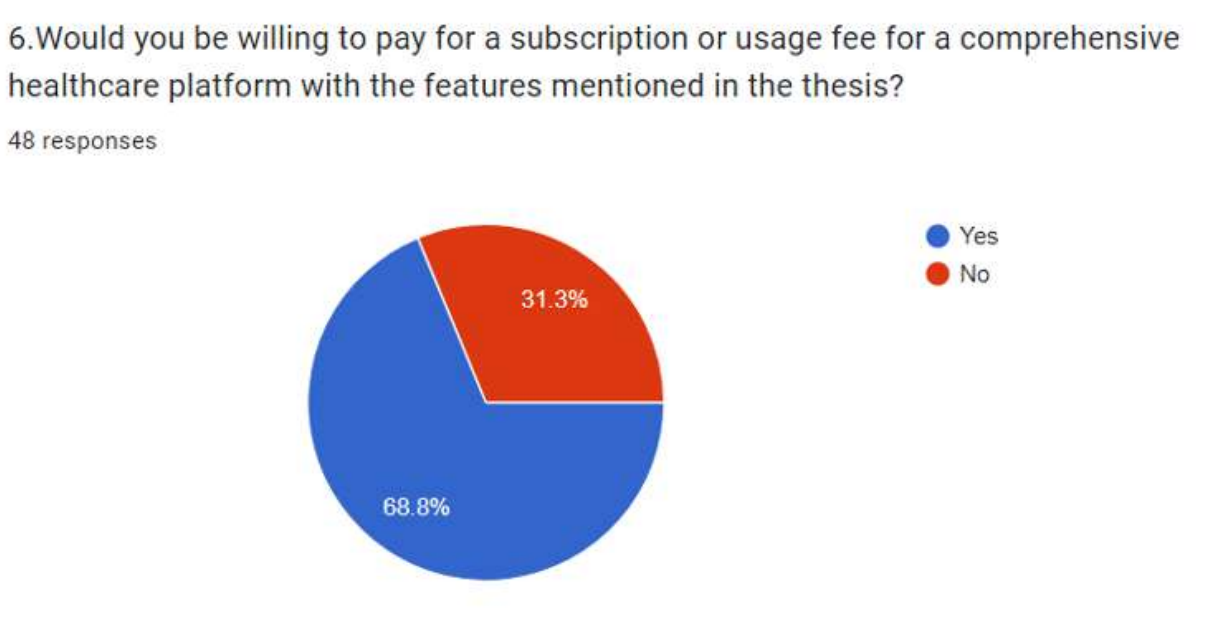}
    \label{fig:model5}
\end{figure}

The resources and instruments available for managing healthcare today frequently leave a lot of people unhappy. This dissatisfaction can be methodically addressed by using the Agile Model. User demands and feedback are regularly included into the process of improving the tools through iterative development. Brief sprints concentrate on improving features and functionality to make sure they meet user requirements. The usability and efficacy of the tools are further improved by frequent testing and feedback loops. Agile Model enables quick adjustments to shifting healthcare needs, resulting in improved management tools. In general, healthcare management resources can adapt to better suit users' requirements and expectations by putting the Agile Model into practice. 

\begin{figure}[H]
    \centering
    \includegraphics[width=0.7\textwidth]{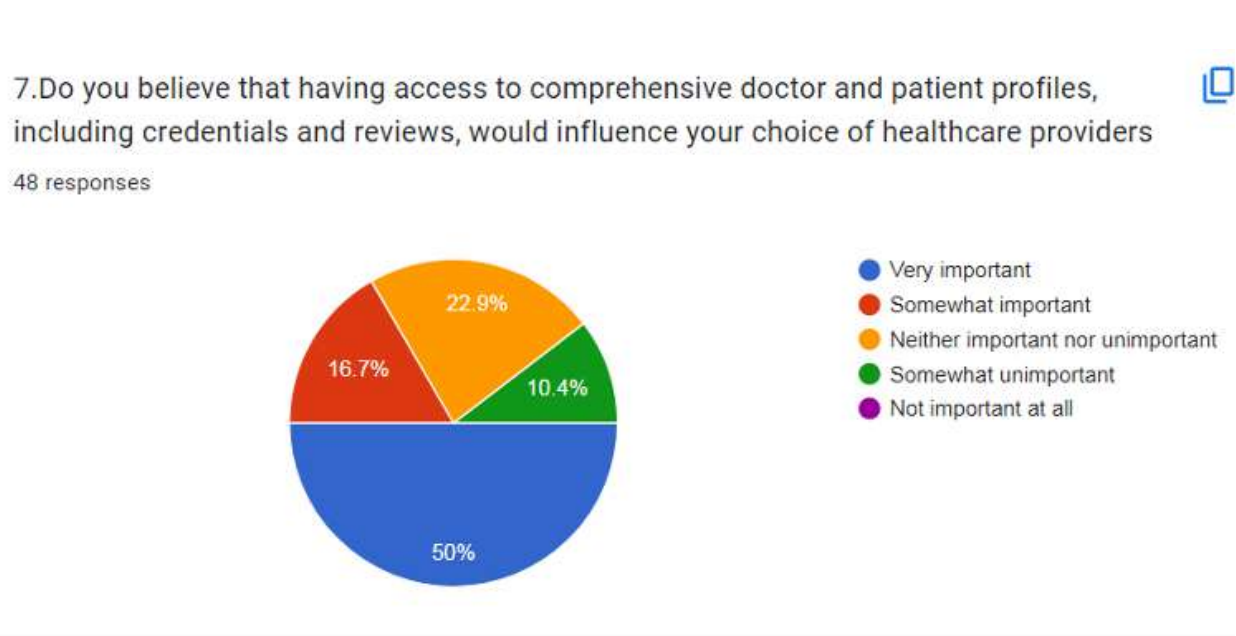}
    \label{fig:model6}
\end{figure}

People are quite likely to use a platform that easily interfaces with hospitals and other healthcare facilities so they can obtain billing information, appointments, and medical data. By utilizing the Agile Model, platform development starts with a comprehensive grasp of user needs, guaranteeing that the platform takes care of important pain points. Sprints are used to prioritize features, enabling quick, incremental development. Iterative testing and continuous feedback loops improve the platform to make sure it flawlessly satisfies user needs. Because of the Agile model emphasis on adaptability, the platform can change to keep up with the norms and practices of healthcare. The platform is positioned to be a useful resource for both patients and healthcare providers because to its frequent updates and user-centric features. The Agile Model ultimately increases the possibility of broad and passionate adoption of the integrated . 

\begin{figure}[H]
    \centering
    \includegraphics[width=0.7\textwidth]{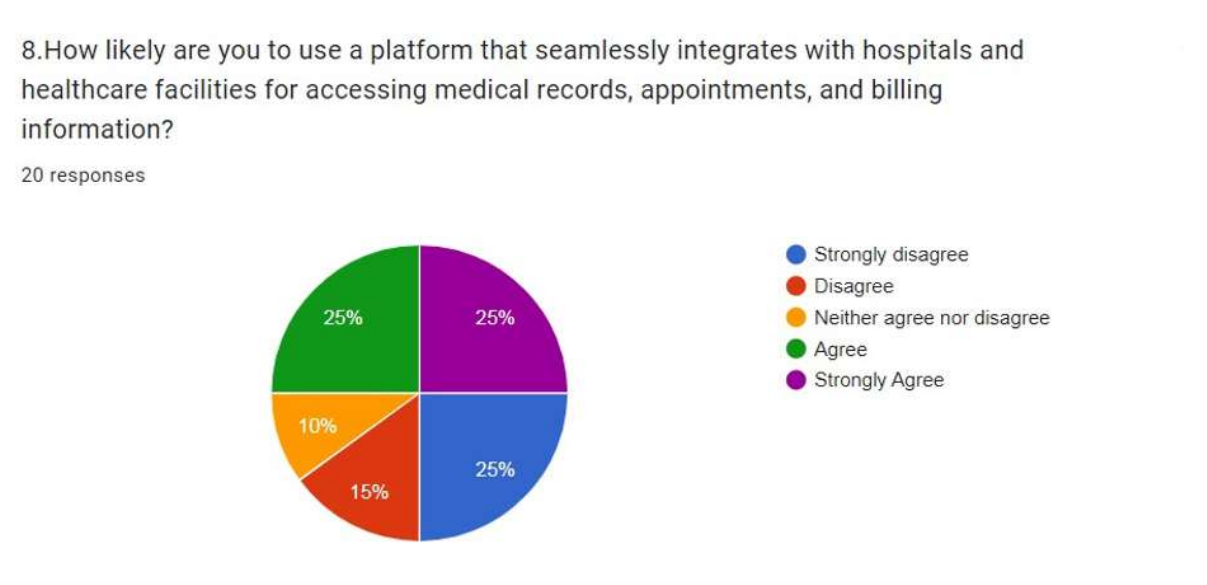}
    \label{fig:model7}
\end{figure}

The resources and instruments available for managing healthcare today frequently leave a lot of people unhappy.
This dissatisfaction can be methodically addressed by using the Agile Model. User demands and feedback are regularly included into the process of improving the tools through iterative development. Brief sprints concentrate on
improving features and functionality to make sure they meet user requirements. The usability and efficacy of the
tools are further improved by frequent testing and feedback loops. Agile Model enables quick adjustments to shifting
healthcare needs, resulting in improved management tools. In general, healthcare management resources can adapt to
better suit users’ requirements and expectations by putting the Agile Model into practice.

\begin{figure}[H]
    \centering
    \includegraphics[width=0.7\textwidth]{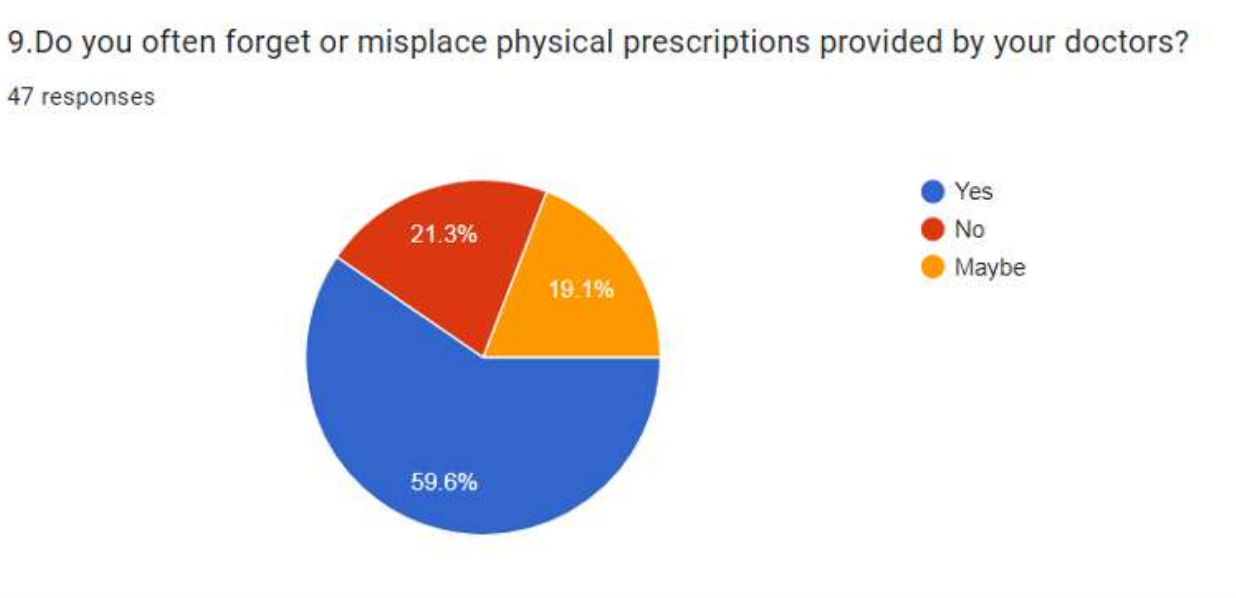}
    \label{fig:model8}
\end{figure}

The resources and instruments available for managing healthcare today frequently leave a lot of people unhappy.
This dissatisfaction can be methodically addressed by using the Agile Model. User demands and feedback are regularly included into the process of improving the tools through iterative development. Brief sprints concentrate on
improving features and functionality to make sure they meet user requirements. The usability and efficacy of the
tools are further improved by frequent testing and feedback loops. Agile Model enables quick adjustments to shifting
healthcare needs, resulting in improved management tools. In general, healthcare management resources can adapt to
better suit users’ requirements and expectations by putting the Agile Model into practice.

\begin{figure}[H]
    \centering
    \includegraphics[width=0.7\textwidth]{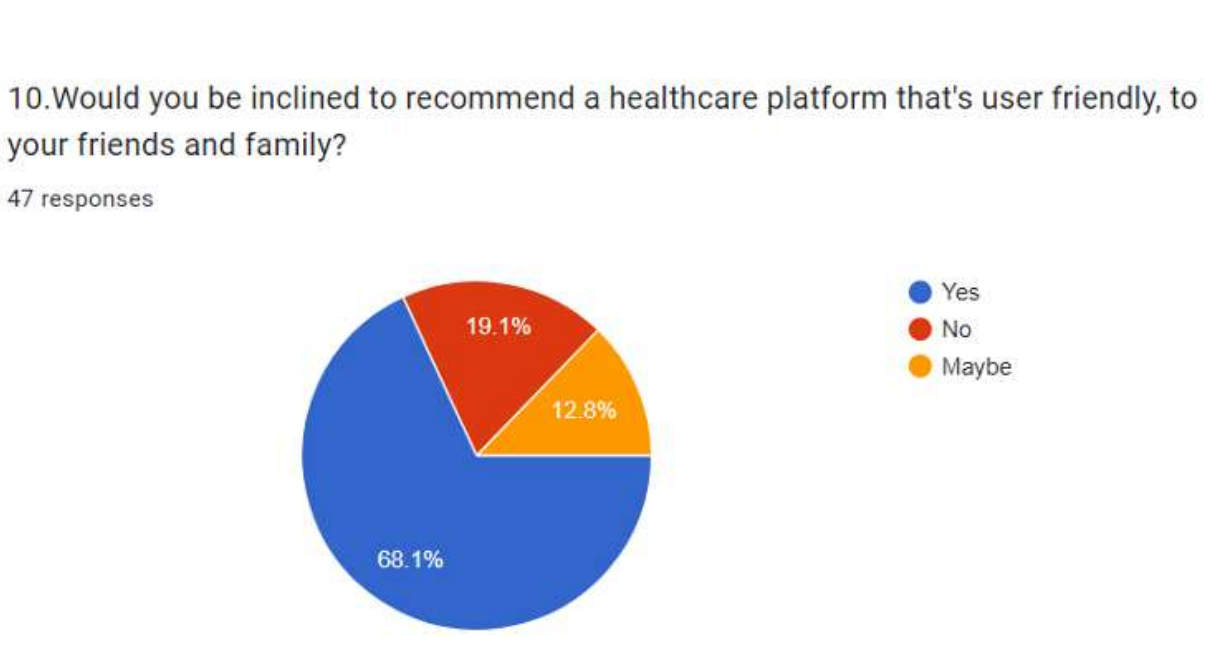}
    \label{fig:model9}
\end{figure}

Human will undoubtedly have a tendency to tell their friends and family about a user-friendly healthcare platform.   The Agile Model ensures that the platform is designed with the user's needs at the forefront. Features are improved through iterative development using ongoing feedback loops. This results in a platform that is very user-friendly, effective, and closely matches user preferences. The site is a useful tool that will probably be shared around social circles due to its user-centric design and seamless connection with healthcare facilities. The platform's attractiveness for recommendations is further enhanced by Agile's regular updates and enhancements. All things considered, the Agile Model's focus on user pleasure and ongoing improvement increases the likelihood that people would enthusiastically recommend such a healthcare platform. 

 \section{Discussion}
 \label{sec:discussion}
This survey investigates the pressing need for agile human resource management systems (HRMS) in the healthcare sector, particularly in response to the increasing complexity of healthcare operations. The implementation of agile methodologies, characterized by iterative improvements, real-time feedback, and flexible adjustments, has demonstrated significant potential for enhancing recruitment, retention, and regulatory compliance processes. The findings of this study indicate that agile approaches contribute to a more responsive and adaptable HR environment, particularly when dealing with the dynamic nature of healthcare personnel management. However, challenges remain, especially in areas concerning cybersecurity and regulatory compliance, which are critical for the sensitive nature of healthcare data. The study identifies that while agile methods offer substantial improvements in operational efficiency, future HRMS implementations must prioritize data security and regulatory adherence to ensure successful outcomes.

 \section{Future Direction}
 \label{sec:futuredirection}
To address these limitations and enhance our understanding of digital Human Resource Management (HRM) systems in healthcare, future research should focus on the scalability and adaptability of such solutions across various hospitals and regions. Building upon the insights gathered in this survey, future research should focus on several key areas. Firstly, there is a need to explore the scalability of agile HRMS across different healthcare settings, from small local clinics to large hospitals and multi-site healthcare networks. This will help determine whether agile frameworks can be universally applied or need customization for specific environments. Additionally, future studies should examine the integration of emerging technologies such as artificial intelligence (AI) and machine learning (ML) to enhance predictive analytics in HR processes, as well as blockchain technology to improve data security and integrity. These technologies have the potential to augment the functionality of agile HRMS, offering new avenues for improving workforce management and compliance. Furthermore, with healthcare regulations continuously evolving, future research should address how agile HRMS can remain compliant while adapting to these changes, ensuring that both legal and operational standards are met.

\section{Conclusion}
\label{sec:conclusion}
The agile human resource management systems (HRMS) has the potential to revolutionize healthcare workforce management by addressing the sector’s complex and evolving challenges. Agile methodologies offer a flexible and responsive framework for managing HR processes in a sector that is characterized by rapid change and high levels of complexity. By promoting continuous improvement, real-time adaptation, and stakeholder involvement, agile HRMS can significantly enhance staff management, improve patient outcomes, and increase organizational efficiency. However, the successful deployment of these systems hinges on addressing key issues such as regulatory compliance and data security. As healthcare continues to evolve, ongoing research and innovation will be essential to ensure that agile HRMS remain relevant and effective, ultimately contributing to the sustainability and growth of healthcare organizations.

\section*{CRediT Authorship Contribution Statement}
\textbf{Syeda Aynul Karim:} Conceptualization, Methodology, Investigation, Writing - Original Draft, Research administration.\newline
\textbf{Md Juniadul Islam:} Methodology, Investigation,  Writing - Review \& Editing,  Resources.

\section*{Declaration of interest’s statement}

There are no competing interests amongst the authors of this article, they declared. As far as the work we have submitted is concerned, we thus declare that we have no competing interests or affiliations.

\section*{The authors acknowledge the resources and supervision provided by ours interest for the Future upcoming research institution "Intelligence System and Machine Learning Lab" (ISML).}

\bibliographystyle{elsarticle-num} 
\bibliography{ref_Unmarked}






\end{document}